\documentclass[aps,prb,floatfix,10pt,twocolumn,superscriptaddress,letterpaper,showpacs]{revtex4}
\usepackage{graphicx,bm,color,amsmath,amssymb,amsfonts,mathrsfs}

\newcommand{\al}{\alpha}
\newcommand{\be}{\beta}
\newcommand{\g}{\gamma}
\newcommand{\de}{\delta}

\newcommand{\ka}{\kappa}

\newcommand{\mi}{\mu}

\newcommand{\p}{\pi}

\newcommand{\W}{\Omega}

\newcommand{\beq}{\begin{equation}}
\newcommand{\eeq}{\end{equation}}
\newcommand{\Beq}{\begin{eqnarray}}
\newcommand{\Eeq}{\end{eqnarray}}
\newcommand{\bml}{\begin{multline}}

\newcommand{\bsp}{\begin{split}}
\newcommand{\esp}{\end{split}}

\renewcommand{\vec}{\overrightarrow}

\newcommand{\nn}{\nonumber}

\newcommand{\bp}{{\bf p}}

\begin{document}
\title{Microscopic theory for a ferromagnetic-nanowire/superconductor heterostructure: Transport, fluctuations
and topological superconductivity}
\author{So Takei}
\affiliation{Condensed Matter Theory Center, The Department of Physics, The University of Maryland, College Park, MD 20742-4111, USA}
\author{Victor Galitski}
\affiliation{Condensed Matter Theory Center, The Department of Physics, The University of Maryland, College Park, MD 20742-4111, USA}
\affiliation{Joint Quantum Institute, The University of Maryland, College Park, MD 20742-4111, USA}
\affiliation{Kavli Institute for Theoretical Physics, University of California Santa Barbara, CA 93106-4030}
\date{\today}

\begin{abstract}
Motivated by the recent experiment of Wang {\em et al.} [Nature Physics {\bf 6}, 389  (2010)],  who  observed a highly unusual transport 
behavior of ferromagnetic Cobalt  nanowires proximity-coupled to superconducting electrodes, we study proximity effect and temperature-dependent transport  in such a mesoscopic hybrid structure. It is assumed that the asymmetry in the tunneling barrier gives rise to the Rashba spin-orbit-coupling in the barrier that enables induced $p$-wave superconductivity in the ferromagnet to exist. We first develop a microscopic theory of Andreev scattering at the spin-orbit-coupled interface, derive a set of self-consistent  boundary conditions, and find an expression for the $p$-wave  minigap in terms of the microscopic parameters of the contact. Second, we study temperature-dependence of the resistance near the superconducting transition and find that it should generally feature a fluctuation-induced peak. The upturn in resistance is related to the suppression of the single-particle density of states due to the formation of fluctuating pairs, whose tunneling is suppressed. In conclusion, we discuss this and related setups involving ferromagnetic nanowires in the context of one-dimensional topological superconductors. It is argued that to realize unpaired end Majorana modes, one does not necessarily need a  half-metallic state, but a partial spin polarization may suffice. Finally, we propose yet another related class of  material systems -- ferromagnetic semiconductor wires coupled to ferromagnetic superconductors -- where direct realization of Kitaev-Majorana model should be especially straightforward. 
\end{abstract}

\maketitle

\section{Introduction}
Coexistence of ferromagnetism and superconductivity is a rare phenomenon, yet there exists unambiguous 
evidence for its occurrence. Superconductivity has been thoroughly investigated in the uranium-based itinerant
ferromagnets\cite{U1,U2,U3,U4} and there have been reports about possible co-existence of these phases 
in the $d$-electron compound  ZrZn$_2$\cite{Zr} and the copper oxide compound RuSr$_2$GdCu$_2$O$_{8-\delta}$ 
(Ru-1212)\cite{RuSr1,RuSr2,RuSr3,RuSr4}. The neutron diffraction measurements in ferromagnetic 
superconductors ErRh$_4$B$_4$\cite{FMS1}, HoMo$_6$S$_8$\cite{FMS2} and HoMo$_6$Se$_8$\cite{FMS3} 
indicate an inhomogeneous magnetic order coexisting with superconductivity. 
More recently, coexistence of these seemingly exclusive states was  also seen in Pb/PbO core/shell 
nanoparticles\cite{nanop}, and in two-dimensional interfaces between perovskite band insulators 
LaAlO$_3$ and SrTiO$_3$\cite{LAOSTO1,LAOSTO2}.

The interplay of superconductivity and ferromagnetism can also be studied in the context of superconducting 
proximity effect, where a superconductor in contact with a normal metal induces superconducting correlations 
in the latter. Due to incompatible spin order conventional superconducting correlations
are known to penetrate negligibly inside a ferromagnet.\cite{buzdinrev} However, a number of recent 
studies have demonstrated an unexpectedly long-ranged proximity effect in mesoscopic superconductor-ferromagnet 
hybrid structures.\cite{SFJ1,SFJ2,SFJ3,SFJ4,wangetal,SFJrev} A subsequent theoretical work showed that 
local inhomogeneity in the magnetization near the interface can induce triplet-pairing inside the ferromagnet, and
that these correlations can account for the observed long-ranged proximity 
effect.\cite{efetovprl} This beautiful theory predicts an exotic odd-frequency (even-momentum) symmetry for the 
triplet component of the condensate, which was originally suggested by Berenzinskii as a possible phase in superfluid 
$^3$He.\cite{berezinskii} Theoretical work on the Josephson coupling between two conventional superconductors
separated by a half-metallic ferromagnet has also investigated the mechanism behind singlet-to-triplet conversion
near a half-metal-superconductor interface.\cite{eschrig1,eschrig2}

Seemingly unrelated to the developments in superconducting proximity effect at the time, Kitaev showed in his 
pioneering work that  Majorana fermion excitations can be localized at the ends of a spinless $p_x+ip_y$ 
superconducting quantum wire.\cite{kitaev} Conceptually, the Kitaev model of a topological superconductor is that
of a fully-polarized ferromagnetic superconductor. However, a direct connection between the model and the existing 
ferromagnetic  itinerant and hybrid superconducting systems has been made only very recently. In some of the recent proposals, superconducting proximity effect plays a key role in realizing topological superconductors which support Majorana fermions on their boundaries or in a vortex 
core.\cite{MF1,MF2,MF3,MF4,MF5,MF6,MF7,MF8,MF9,MF10,MF11,MF12,MF13} 

On the experimental front, Ref. \onlinecite{wangetal} has already shown that a ferromagnetic cobalt (Co) quantum wire
can be made superconducting by placing it in contact with a conventional superconductor. Proximity-induced superconductivity
was observed in the wire over a distance of a few hundred nanometers. The pairing symmetry inside the 
wire is likely to be a spin-triplet state since singlet correlations should not survive inside the ferromagnet over a distance exceeding
a few nanometers. While the odd-frequency superconductivity indeed remains a viable explanation of the long-range proximity effect, 
an alternative scenario in clean wires involves $p$-wave superconductivity induced by a spin-orbit coupled interface.\cite{MF8} 
The latter would imply that an experiment of the kind in Ref. \onlinecite{wangetal} is very close to realizing the topological 
superconductor\cite{kitaev}, and may host the sought-after Majorana fermions at the two ends of the wire or inside the wire, wherever
the parity of the number of occupied sub-bands changes.

With its broader physical relevance aside, the experimental work of Ref. \onlinecite{wangetal} has made very interesting 
observations in the context of superconducting proximity effect in ferromagnetic systems. The work systematically studies
the resistance of ferromagnetic single-crystal cobalt nanowires sandwiched between two superconducting electrodes. They observe 
proximity-induced superconductivity in the nanowire over a distance of order 500nm, which is orders of magnitude 
longer than the coherence length expected for conventional superconducting correlations inside a ferromagnet. For some
wires, transition to superconductivity is preempted by a large and sharp resistance peak near the transition temperature of
the electrodes. The resistance peak disappears when the
Co wire is replaced by a gold wire, indicating an intimate connection between it and the ferromagnetism of
the wire.

Motivated by these observations, we first theoretically study the physics of an $s$-wave-superconductor-ferromagnet tunnel 
junction. We assume that the asymmetry in the tunneling barrier gives rise to the Rashba spin-orbit coupling inside the tunnel 
barrier and enables $p$-wave superconductivity to be induced inside the ferromagnet.\cite{MF8} We derive a complete 
set of relevant boundary conditions, and an expression for the $p$-wave minigap is found in terms of the microscopic parameters 
of the contact. In the second part of the work, we investigate in detail how the anomalous resistance peak can be explained based on the 
theory of superconducting fluctuation corrections to conductivity.\cite{varlarrev,varlamovetal} We discuss a possible
faithful model for the cobalt nanowire in the vicinity of the deposited W electrodes, develop a microscopic theory for the
superconducting fluctuations in the W electrodes, and study how these fluctuations influence the resistance of the cobalt
wire. Finally, we discuss in more detail the possible relevance of the experiment to topological superconductivity and propose a related
hybrid system, which provides in our opinion the simplest physical realization of topological superconductivity and localized Majorana modes.

We emphasize that the Co wire throughout the work is treated mostly in the ballistic/ballistic-to-diffusive crossover limit. As we outline in later parts of the paper, the treatment of the wire in this limit is reasonably justified based on experimental data\cite{wangetal} and past studies on Co band structure\cite{Coband1, Coband2,Coband3,Coband}, although the information available to us is insufficient to conclusively determine the nature of proximity-induced superconductivity in the hybrid system.  As we have mentioned above, long-ranged \textit{odd-frequency} $p$-wave proximity effect, extensively studied in diffusive ferromagnets \cite{efetovprl,SFJrev}, remains another realistic scenario.   Here, we are exploring the possibility of a  long-ranged \textit{spatially-odd} $p$-wave proximity effect induced via interfacial spin-orbit coupling, which 
remains a viable physical scenario when the ferromagnet is in the clean limit. Furthermore, as argued below, this latter mechanism may survive  even in disordered systems due to mesoscopic fluctuation effects. 

The paper is organized as follows. Sec. \ref{experiment} summarizes the main findings of the experiment in Ref. \onlinecite{wangetal},
with a particular emphasis on the critical resistance peak. The section also lays out in detail an estimate for the mean-free path of the Co wire 
and a justification for considering the clean limit. A  microscopic theory for Andreev scattering at a ferromagnet-superconductor
interface is presented in Sec. \ref{andreev}. In Sec. \ref{sec:SCfluc}, the superconducting fluctuation theory is applied to the
W electrodes. Following a qualitative discussion on how superconducting fluctuations in the W can lead to resistance peaks
in Co transport, a detailed microscopic theory for the corrections to resistance from these fluctuations is developed in
Sec. \ref{micro}. A discussion of topological superconductivity is presented in Sec.~\ref{TopSC}. 

\section{Summary of the Penn State Experiment}
\label{experiment}
\begin{center}
\begin{figure}[t]
\includegraphics*[scale=0.6]{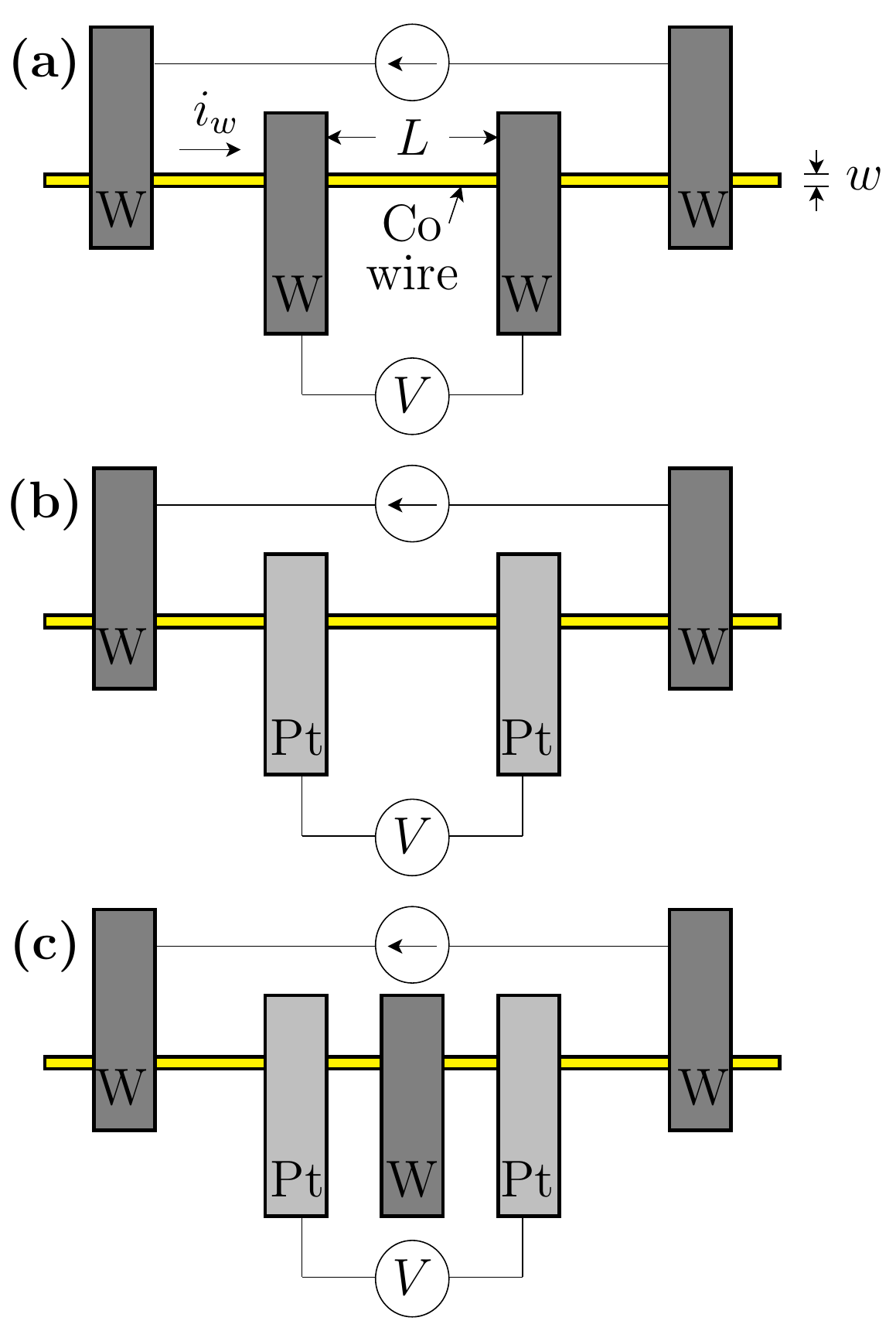}
\caption{\label{fig:sys} A cartoon representation of the various experimental setups used in Ref. \onlinecite{wangetal}. 
Three main setups are considered: (a) W current and voltage electrodes; (b) W current electrodes and Pt voltage
electrodes; and (c) same as (b) with an additional electrically isolated W strip between the Pt electrodes. Length of
the wire $L$ is determined by the distance between the inner edges of the voltage electrodes. The width (or diameter) of the
Co wire is denoted by $w$.}
\end{figure}
\end{center}

Ref. \onlinecite{wangetal} reports observations of long-ranged proximity effect in single-crystal ferromagnetic 
Co nanowires.  Resistance of the wire is studied using a four-probe setup with all four electrodes 
made from superconducting tungsten (W) [see Fig. \ref{fig:sys}(a)] or with a combination of superconducting W and
non-superconducting platinum (Pt) electrodes [see Fig. \ref{fig:sys}(b)]. While the outer electrodes are used to pass
current $i_w$ through the wire, the inner electrodes are used to measure the potential difference across a
length $L$ of the wire. With superconducting voltage electrodes [Fig. \ref{fig:sys}(a)] Co wires become fully 
superconducting below the electrode transition temperature for $L=600$nm, and show substantial resistance drop 
for $L=1.5\mu$m. However, such
resistance drop was not observed when the superconducting electrodes are replaced by non-superconducting Pt. 
However, when an additional electrically isolated W strip is deposited between the Pt 
electrodes (see Fig. \ref{fig:sys}(c)), proximity-induced superconducting properties are restored in the Co wire.

\subsection{Critical resistance peak}
\begin{center}
\begin{figure}[t]
\includegraphics*[scale=0.55]{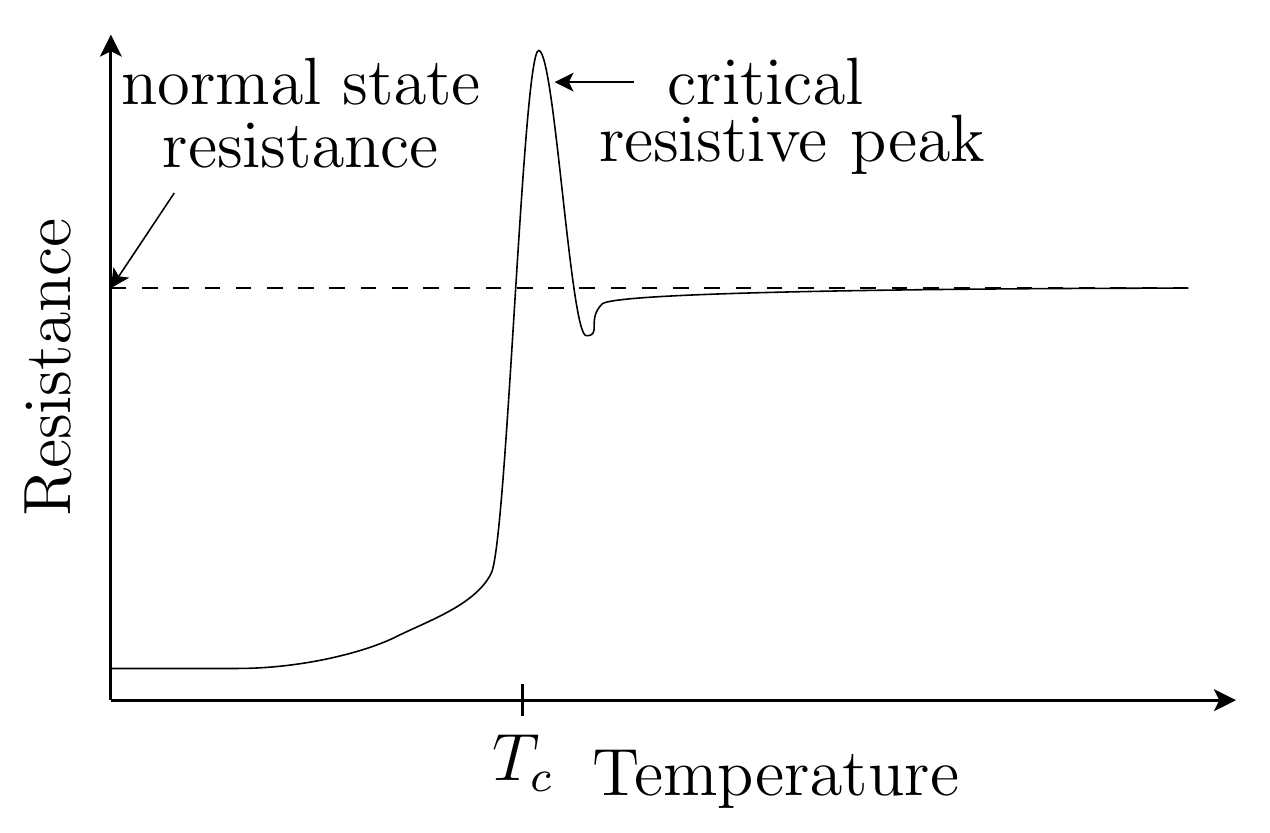}
\caption{\label{fig:RvsT} A cartoon plot of the experimentally observed Co wire resistance as a function of 
temperature. A critical resistance peak is observed near the transition temperature $T_c$ of the W electrodes. A precipitous
drop in the resistance is then observed below $T_c$.}
\end{figure}
\end{center}

The most interesting observation in the experiment is that the resistance shows a sharp peak 
as a function of temperature as it approaches the 
superconducting transition temperature of the W electrodes. A cartoon picture of a typical resistance versus
temperature curve is shown in Fig. \ref{fig:RvsT} to illustrate this result. The transition temperature of the electrodes
is estimated to be between $T_c=4.4$-5K. This resistance peak is very large and constitutes 25-100\% 
of the normal state resistance depending on the wire width. Intriguingly, the peak disappears when
the Co wire is replaced by a paramagnetic gold wire, and thus seems to be a consequence of the
ferromagnetism of the Co wire. The large resistance peak also disappears when the W electrodes are
replaced by Pt electrodes [see Fig. \ref{fig:sys}(b)], but is restored once an electrically isolated W strip is
deposited [see Fig. \ref{fig:sys}(c)]. The large resistance upturn thus seems to require only that a superconducting 
strip is in contact with the nanowire, and does not require it to be either a current or a voltage electrode.

\subsection{Mean-free path}
\label{meanfree}
When the extra W strip is deposited on the wire [see Fig. \ref{fig:sys}(c)] 
data show that the wire's normal state resistance increases by approximately
55$\W$. This amounts to a nearly 50\% increase from its normal state resistance of $\approx$126$\W$ prior to the 
deposition [see Fig. \ref{fig:sys}(b)]. This suggests that the deposition process could have strongly modified the property 
of the wire in the vicinity of the contact. Returning to the setup in Fig. \ref{fig:sys}(a) the
experimental data show total normal state wire resistance of approximately 145$\W$.
The resistance of the Co wire arising from regions unaffected by the W strips can then be approximated as
$145\W-2\cdot 55\W=35\W$. Using the quoted distance between the voltage electrodes (i.e. $L=1.5\mu$m) and the 
wire width (i.e. $w=36$nm) the corresponding resistivity is $\rho\approx 23$n$\W\cdot$m. An estimate for 
Co's Fermi velocity, based on critical current oscillations observed in a Nb/Co/Nb Josephson junction, 
gives $v_F\approx 280$km/s.\cite{Covel1} From this the density for electrons with a single spin
projection can be estimated as 
\beq
n=x^3\frac{m_e^3v_F^3}{6\p^2\hbar^3}\approx x^3\cdot2.4\times 10^{26}\mbox{m}^{-3},
\eeq
where $x=m^*/m_e$ is the Co effective mass ratio with $m^*$ being the effective mass of the  
Co electrons and $m_e$ the
mass of the electron. The Drude formula then gives a mean-free path estimate of
\beq
\label{MFPath}
\ell=\frac{xm_ev_F}{ne^2\rho}\approx\frac{1.8\mi\mbox{m}}{x^2}.
\eeq

Cores of the Co nanowires studied in Ref. \onlinecite{wangetal} show a hexagonal close-packed zone pattern and the 
wires have a [0001] growth direction. Presumably, the electron transport through the wire is then along the $c$-axis 
direction. For our estimate of the Co effective mass $x$, we use results based on band structure calculations for Co
cyclotron mass.\cite{Coband} In the relevant direction of transport, we have $x\approx 3$. 
Inserting this value into (\ref{MFPath}) the mean-free path is estimated to be $\ell\approx 200$nm, which is 
the same order of magnitude as the observed coherence length in the Co wire. Within this scenario, regions of the 
Co wire unaffected by the W electrodes can be considered in the clean limit. 

We note, however, that the above estimate for the mean-free path relies crucially on our estimate for the resistance of 
the Co wire unaffected by the W electrodes. Furthermore, the estimate also depends sensitively on the values 
used for $v_F$ and $x$, which we do not know with certainty. Due to Co's complicated band structure, the values for 
$v_F$ and $x$ strongly depend on the direction of transport with respect to the crystallographic axes and on the 
Fermi surface which gives the dominant contribution to transport.
For instance, for $v_F\sim 10^6$m/s (quoted in Ref. \onlinecite{wangetal}) 
we obtain a much shorter mean-free path of $\ell\approx 16$nm, implying 
that the wire is in the diffusive limit. Co having multiple subbands with varying $x$ also implies that some subbands are
less affected by disorder than others. If all bands participating in transport indeed have short mean-free paths (of order a
few nanometers),  the likely mechanism for proximity effect is the spatially even, odd-frequency pairing.\cite{SFJrev,efetovprl} 
However, based on experimental data in Ref. \onlinecite{wangetal} and on Co band structure and 
de-Haas-van Alphen studies relevant for electron transport in the $c$-axis direction\cite{Coband}, we have shown above that 
there is reasonable justification to model the Co wire in the clean limit, and the spatially odd $p$-wave pairing is a viable mechanism
for the observed proximity effect. Furthermore, mesoscopic fluctuation effects may further enable remnants of $p$-wave 
superconductivity to persist well into the diffusive limit.\cite{pwpe} In this work, we pursue this scenario of a Co wire with 
proximity-induced $p$-wave superconductivity, and purport that it provides a natural explanation for the resistance peak.

\section{Andreev reflection at a FM-SC interface}
\label{andreev}
\begin{center}
\begin{figure}[t]
\includegraphics*[scale=0.38]{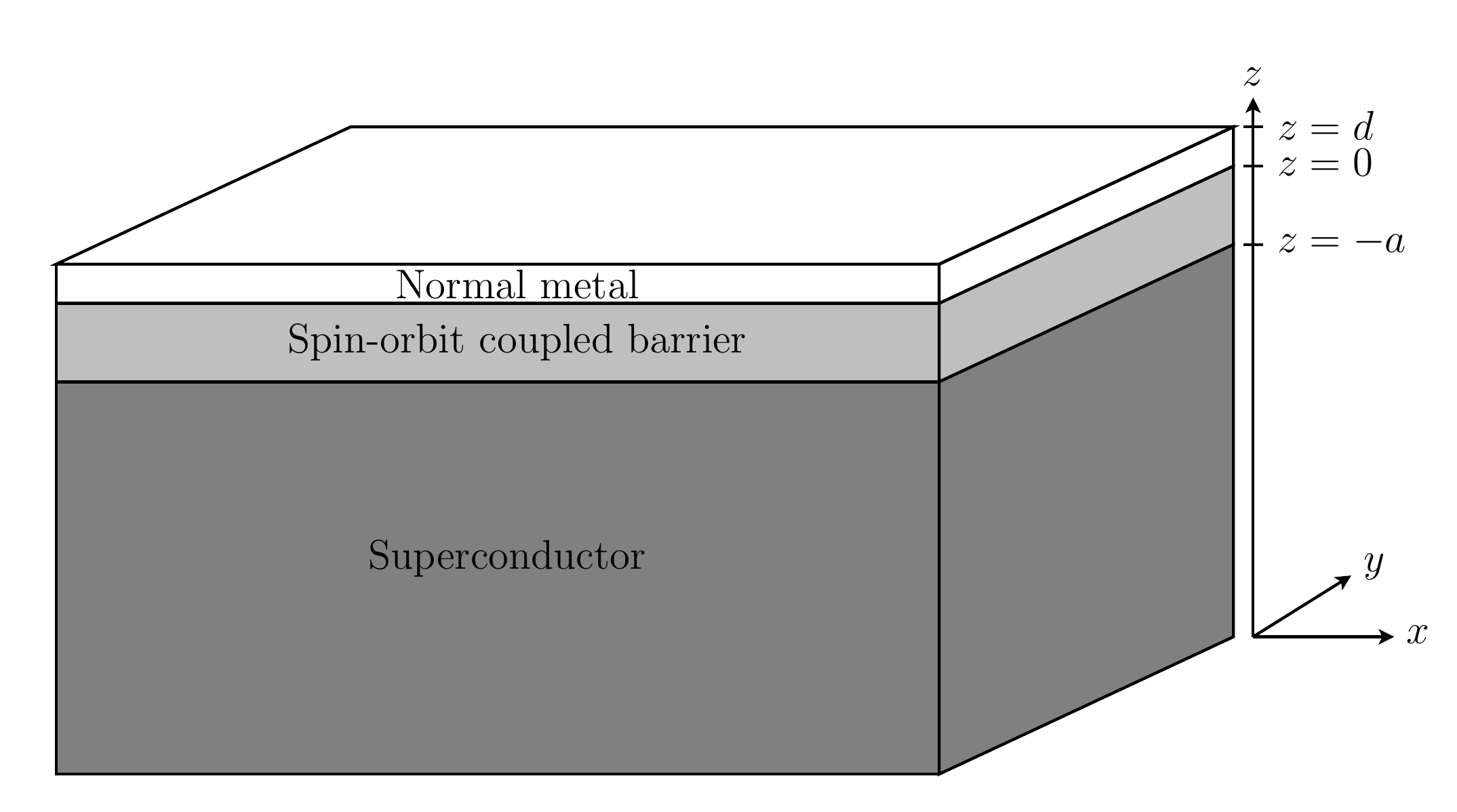}
\caption{\label{fig:sys2} The proximity tunnel junction considered. The ferromagnetic normal metal, the tunnel barrier with spin-orbit
coupling and the superconductor occupy $0\le z\le d$, $-a<z<0$ and $z\le-a$, respectively.}
\end{figure}
\end{center}

We consider a tunnel junction consisting of a ferromagnetic normal metal at $0\le z<d$, a tunneling barrier at 
$-a< z<0$, and a superconductor at $z \le -a$ (see Fig. \ref{fig:sys2}). 
At the moment, we assume translational symmetry in the direction(s) parallel
to the interfaces. We will work with a planar junction, but the results will also describe proximity effect in a ferromagnetic nanowire 
by simply ``removing'' one of the two transverse directions or by choosing a specific transverse sub-band. The derivation below follows closely that of Refs. \onlinecite{ProxEff1} 
and \onlinecite{ProxEff2}, but is generalized to include spin-orbit-coupling effects in the barrier. 

\subsection{Equations of motion}
The electron field operators in the ferromagnet $\psi^{(F)}$, tunnel barrier $\psi^{(B)}$ and superconductor $\psi^{(S)}$ satisfy 
the following equations of motion. In the normal region ($0\le z<d$),
\begin{equation}
\label{FEq}
\left[ \left( \varepsilon-\xi_{\bf p}^{(F)}  + \frac{\hbar^2 \partial_z^2} {2 m_F} \right) \delta_{\alpha \beta} 
- ({\bm h} \cdot {\bm \sigma}_{\alpha \beta})\right]\psi_{\beta, {\bf p}}^{(F)} (z) = 0;
\end{equation}
in the barrier ($-a< z<0$),
\begin{multline}
\label{BEq}
\left[\left( \varepsilon-U_0-\frac{{\bf p}^2 - \hbar^2 \partial_z^2} {2 m_B} \right) \delta_{\alpha \beta}  
- \alpha_R [{\bm p} \times {\bm \sigma}_{\alpha\beta}]_z  \right]\\\times\psi_{\beta, {\bf p}}^{(B)} (z) = 0;
\end{multline}
and in the superconductor ($z\le -a$),
\begin{equation}
\label{SEq}
\ \left( \varepsilon- \xi_{\bf p}^{(S)}  + \frac{\hbar^2 \partial_z^2} {2 m_S} \right) \psi_{\alpha, {\bf p}}^{(S)} (z) -
\Delta_{\alpha \beta}  \psi_{\beta, -{\bf p}}^{(S) \dagger} (z) = 0.
\end{equation}
Here, ${\bf p} = (p_x,p_y)$ is the two-dimensional momentum parallel to the interface, 
$\hat{\bm \sigma} = \left( \hat{\sigma}_x,\hat{\sigma}_y, \hat{\sigma}_z\right)$ is the vector of Pauli 
matrices acting in spin space,  $\xi_{\bf p}^{(F)}  = {\bf p}^2/2 m_F - \mu_F$ and  $\xi_{\bf p}^{(S)}  
= {\bf p}^2/2 m_S - \mu_S$ are the electronic spectra in the ferromagnet and superconductor, 
$m_F$, $m_S$ and $m_B$ are the effective masses in the corresponding region and $\mu_F$ and 
$\mu_S$ are the Fermi levels in the ferromagnet and superconductor. $\al$ and $\be$ here label the
spin projections. We assume that the ferromagnet 
and superconductor are separated by a tunnel barrier of height $U_0$ that, due to mirror asymmetry with respect to 
reflections relative to the $z = -a/2$ plane, contains Rashba spin-orbit coupling characterized by 
parameter $\alpha_R$. The ferromagnet is modeled as a Fermi gas in a Zeeman field ${\bm h}$.
Finally, the superconductor is assumed to be of $s$-wave type and, consequently, the mean-field order 
parameter is momentum independent and reads $\Delta_{\alpha \beta}  = \Delta (i \sigma_{y,\alpha \beta})$.

Although the barrier is spin-orbit coupled, this coupling will not lead to any spin Hall effects at the boundaries 
$z=0$ and $z= -a$. Consequently, there is no accumulation of spin and the basic ballistic boundary conditions 
are identical to those in the theory of a conventional tunnel junction, namely:
\begin{equation}
\label{b1}
\psi_{\alpha, {\bf p}}^{(F)} (d) = 0,
\end{equation} 
\begin{equation}
\label{b2}
\psi_{\alpha, {\bf p}}^{(F)} (0) =  \psi_{\alpha, {\bf p}}^{(B)} (0),\qquad
 \psi_{\alpha, {\bf p}}^{(B)} (-a) = \psi_{\alpha, {\bf p}}^{(S)} (-a),
\end{equation}
\begin{equation}
\label{b3}
\frac{1}{m_F} \partial_z \psi_{\alpha, {\bf p}}^{(F)} (0) =  \frac{1}{m_B} \partial_z  \psi_{\alpha, {\bf p}}^{(B)} (0),
\end{equation}
and
\begin{equation}
\label{b4}
\frac{1}{m_B} \partial_z \psi_{\alpha, {\bf p}}^{(B)} (-a) =  \frac{1}{m_S} \partial_z  \psi_{\alpha, {\bf p}}^{(S)} (-a).
\end{equation}
The full set of equations (\ref{FEq})-(\ref{SEq}) is formulated for electron field operators acting in Fock space. 
However, since these equations and boundary conditions  (\ref{b1})-(\ref{b4}) are all linear, the operator nature 
of the unknown functions is not germane and we may simply approach the problem as in single-particle 
quantum mechanics. The problem is then conceptually very simple and reduces to solving second-order differential 
equations albeit with non-trivial boundary conditions. This can be done exactly to a large degree. 

\subsection{Deriving closed boundary conditions \\ for the ferromagnet}

\subsubsection{Tunnel barrier}
\label{ss:tb}
Let us look for a solution in the tunnel barrier in the following form, $\psi_{\alpha, {\bf p}}^{(B)} (z) = Z(z) {\cal P}_\alpha({\bf p})$,
where the transverse wave function $Z(z)$ satisfies  
\beq
-{\hbar^2 \over 2m_B} Z'' + [U_0 - \varepsilon + \varepsilon^{R}({\bf p})  ] Z = 0,
\eeq
and the planar wave function ${\cal P}_\alpha({\bf p})$ is a standard wave function of the Rashba problem,
\begin{equation}
\label{Rashbasol}
{\cal P}^\pm ({\bf p}) = {1 \over \sqrt{2}} {\pm 1 \choose - i e^{i \gamma_{\bf p}}},
\end{equation}
describing the helicity eigenstates with eigenvalues, $\varepsilon^{R}_\pm({\bf p}) = {{\bf p}^2 \over 2m_B} \pm \alpha_R |{\bf p}|$. 
The angle $\gamma_{\bf p}$ is defined via $ {\bf p} = p \left( \cos{\gamma_{\bf p}}, \sin{\gamma_{\bf p}} \right)$.
The general solution in the barrier then reads
\begin{eqnarray}
\label{gensolbar}
\psi_{\alpha, {\bf p}}^{(B)} (z) = && \left[ C_{++} e^{q_+ z} +  C_{+-} e^{-q_+ z} \right] {\cal P}^+_\alpha ({\bf p}) \nonumber \\
&&\!\!\!\!\! + \left[ C_{-+} e^{q_- z} +  C_{--} e^{-q_- z} \right] {\cal P}^-_\alpha ({\bf p}),
\end{eqnarray}
where $q_\pm = \sqrt{ {2 m_B \over \hbar^2} [U_0 - \varepsilon + \varepsilon^{R}_\pm({\bf p})  ] }$. To guarantee real
$q_\pm$ we assume that the spin-orbit interaction scale is smaller than the barrier height.

We now use (\ref{b2}) to determine the relations between the coefficients in (\ref{gensolbar}) and the values of the 
wave function on the superconductor and ferromagnet boundaries. We find
\begin{equation}
\label{Css}
C_{ll'} = {l'  \over 4 \sinh(q_l a)} \left( {\cal F}^l e^{l' q_l a} - {\cal S}^l \right), 
\end{equation}
where $l,l' =\pm$ and
\begin{align}
\label{F+-}
{\cal F}^\pm &= i \psi_{\downarrow, {\bf p}}^{(F)}(0) e^{-i \gamma_{\bf p}} \pm  \psi_{\uparrow, {\bf p}}^{(F)}(0);\\
{\cal S}^\pm &= i \psi_{\downarrow, {\bf p}}^{(S)}(-a) e^{-i \gamma_{\bf p}} \pm  \psi_{\uparrow, {\bf p}}^{(S)}(-a).
\end{align}

\subsubsection{Superconductor}
We now match the tunnel-barrier solution (\ref{gensolbar}) with the superconducting solution at the barrier-superconductor
interface using boundary condition (\ref{b4}). We then find
\begin{equation}
\label{supbound}
{m_B \over m_S} \psi_{\alpha, {\bf p}}^{(S)}(-a) = T_{\bp,\alpha \beta}  \psi_{\beta, {\bf p}}^{(F)}(0) - R_{\bp,\alpha \beta}  \psi_{\beta, {\bf p}}^{(S)}(-a).
\end{equation}
The matrices in (\ref{supbound}), which play an important role in the analysis of the Andreev scattering problem, are given by
\begin{align}
\label{T}
\hat{T}_\bp &= \begin{pmatrix} \kappa_t & i \delta\kappa_t e^{-i \gamma_{\bf p}} \\  -i \delta\kappa_t e^{i \gamma_{\bf p}} & \kappa_t \end{pmatrix}\\
\label{R}
\hat{R}_\bp&= \begin{pmatrix} \kappa & i \delta\kappa e^{-i \gamma_{\bf p}} \\  -i \delta\kappa e^{i \gamma_{\bf p}} & \kappa \end{pmatrix}.
\end{align}
Here, we have defined
\begin{eqnarray}
\label{kappat}
\kappa_t = {q_+ \over 2 \sinh(q_+ a)} + {q_- \over 2 \sinh(q_- a)} \\
 \delta \kappa_t = {q_+ \over 2 \sinh(q_+ a)} - {q_- \over 2 \sinh(q_- a)}
\end{eqnarray}
and
\begin{eqnarray}
\label{kappa}
\kappa = {q_+ \over 2 \tanh(q_+ a)} + {q_- \over 2 \tanh(q_- a)} \\
 \delta \kappa = {q_+ \over 2 \tanh(q_+ a)} - {q_- \over 2 \tanh(q_- a)},
\end{eqnarray}
with $q_\pm^{-1}$ being the penetration length of a particle with a positive/negative chirality inside the barrier, and $a$ 
is the barrier width. As the width of the barrier grows, the ``tunneling'' boundary coefficients 
decay exponentially, $\lim_{a \to \infty} \kappa_t = 0$. As expected, we see then that the coupling between the superconductor and the 
ferromagnet disappears, as does the proximity effect. Furthermore, if we ``turn off'' the spin-orbit coupling in the barrier 
(i.e. setting $\alpha_R = 0$, so that $q_+ =q_-$), we see that the spin-mixing terms vanish, $\delta \kappa = \delta \kappa_t = 0$, 
and we recover the standard boundary conditions for the proximity effect.

To describe the superconductor [see (\ref{SEq})] we introduce a four-component state-vector in the combined spin-Nambu space,
$$\vec{\Psi}_{\bf p}(z) = \left(  \psi_{\uparrow, {\bf p}}^{(S)}(z), \psi_{\downarrow, {\bf p}}^{(S)}(z), \psi_{\uparrow, -{\bf p}}^{(S)*}(z), 
\psi_{\downarrow, -{\bf p}}^{(S)*}(z) \right)^T.$$ 
The Pauli matrices in Nambu space will be denoted by $\hat\tau_x$, $\hat\tau_y$ and $\hat\tau_z$, and the unit matrix by
$\hat\tau_0$. $4\times4$ matrices in the spin-Nambu space will be denoted by an inverse hat, for instance,
\begin{equation}
\label{BCNambu}
\check{T}_{\bf p} = \begin{pmatrix}  \hat{T}_{\bf p} & 0 \\  0&  \hat{T}_{-\bf p}^* \end{pmatrix},\qquad
\check{R}_{\bf p} = \begin{pmatrix} \hat{R}_{\bf p} & 0 \\  0&  \hat{R}_{-\bf p}^* \end{pmatrix}  ,
\end{equation}
where the asterisk denotes complex conjugation. Note that $\gamma_{-\bf p} = \gamma_{\bf p} + \pi$, therefore, 
$\exp{(i\gamma_{-{\bf p}})} = - \exp{(i\gamma_{\bf p})}$, as it appears in $\hat{T}_{-\bf p}^*$ and $\hat{R}_{-\bf p}^*$.

Using these notations, we can write the boundary conditions for the superconductor in the following compact form, 
${m_B \over m_S} \partial_z \vec{\Psi}_{\bf p}^{(S)}(-a) = \check{T}_{\bf p}  \vec{\Psi}_{\bf p}^{(F)}(0) - \check{R}_{\bf p}  
\vec{\Psi}_{\bf p}^{(S)}(-a)$. We proceed by first incorporating the right side of this boundary condition 
into the Bogoliubov-de Gennes equation for a superconductor, which we symbolically write in the form,\cite{ProxEff2}
\begin{multline}
\label{shiftBCBdG}
\check{\tilde G}_{\bf p}^{-1} \vec{\Psi}_{\bf p}^{(S)}(z) \\ = - {\hbar^2 \over 2 m_B} \delta(z + a) 
\left[   \check{T}_{\bf p}  \vec{\Psi}_{\bf p}^{(F)}(0) - \check{R}_{\bf p}  \vec{\Psi}_{\bf p}^{(S)} (-a)\right].
\end{multline}
Here, $\check{\tilde G}_{\bf p}(z,z')$ is the Green function of the Bogoliubov-de Gennes equation for a bulk superconductor 
in the half-space, $z<-a$, which satisfies the von Neumann boundary conditions, $\partial_z \check{\tilde G}_{\bf p}(z,z')\Bigl|_{z=-a} = 0$. 
It can be expressed in terms of the Green function of an infinite bulk $s$-wave superconductor by the method of mirror images and 
reads,
\begin{equation}
\label{mirror}
\check{\tilde G}_{\bf p}(z,z') = \check{G}_{\bf p}(z - z') + \check{G}_{\bf p}(z + z'+2a),
\end{equation} 
where 
\beq
\check{G}_{\bf p} =  \begin{pmatrix}  \hat{G}_{\bf p} &\hat{F}_{-{\bf p}}^* \\ \hat{F}_{\bf p} & \hat{G}_{-{\bf p}}^* \end{pmatrix}
\eeq 
and $ \hat{G}_{\bf p}$ and $ \hat{F}_{\bf p}$ are the normal and Gor'kov Green functions, respectively. Consequently, the solution to the 
Bogoliubov-de Gennes equation (\ref{shiftBCBdG}) is derived by convoluting the Green function with the boundary term on the 
right side of (\ref{shiftBCBdG}), which is simple due to the delta-function in the latter,
\begin{equation}
\label{solBdG}
 \vec{\Psi}_{\bf p}^{(S)}(z)  = - {\hbar^2 \over m_B} \check{G}_{\bf p}(z + a)  \left[   \check{T}_{\bf p}  \vec{\Psi}_{\bf p}^{(F)}(0) 
 - \check{R}_{\bf p}  \vec{\Psi}_{\bf p}^{(S)} (-a)\right].
 \end{equation}

For the purpose of understanding induced superconductivity in the ferromagnet we only need to know the boundary value of 
the superconducting wave function at $z=-a$. From (\ref{solBdG}) we obtain
\begin{equation}
\label{sol(-a)}
 \vec{\Psi}_{\bf p}^{(S)}(-a)  = - {\hbar^2 \over m_B} {1 \over 1 - (\hbar^2/m_B) \check{g}_{\bf p}  \check{R}_{\bf p} } \check{g}_{\bf p}   \check{T}_{\bf p}  \vec{\Psi}_{\bf p}^{(F)}(0),
 \end{equation}
 where 
\begin{equation}
\label{g}
 \check{g}_{\bf p} = \lim_{z\to -a} \check{G}_{\bf p}(z + a) = \int {d p_z \over 2 \pi}  \check{G}(p_x,p_y,p_z).
\end{equation}
We note here that (\ref{sol(-a)}) is {\em exact} and that we have not used any properties of the system at $z>0$ up to this point. 
Therefore, the equation above is applicable to any such junction with any normal or superconducting material at $z>0$ 
(the only constraint is the continuity of derivatives at the $z=0$ boundary, which may be violated if there is a spin Hall effect 
present. In this case, however, an alternative set of boundary conditions can be derived).

\subsubsection{Ferromagnet}
The expression in (\ref{sol(-a)}), which represents a useful technical result of the paper, allows one to formulate a closed 
problem on the (ferromagnetic) normal side. Let us write the Schr{\"o}dinger equation (\ref{FEq}) as
\begin{equation}
\label{FEq2}
\check{G}^{(F)}_{\bf p}\!\!{\phantom{|}}^{-1} \circ \vec{\Psi}_{\bf p}^{(F)}(z) = 0,
\end{equation}
where we have extended the equation into the Nambu space. The propagator $\check{G}^{(F)}_{\bf p}(z,z')$ describes a particle of the 
magnetized Fermi liquid in the shell $0 \le z \le d$. Apart from the trivial boundary condition (\ref{b1}) at the hard wall, 
(\ref{b3}) together with the solution of Sec. \ref{ss:tb} and (\ref{sol(-a)}) give rise to the following constraint,
\begin{equation}
\label{bc(0)}
{m_B \over m_F} \partial_z \vec{\Psi}_{\bf p}^{(F)}(0)  =\left[ \check{R}_{\bf p} +  \check{T}_{\bf p} 
{{\hbar^2 / m_B} \over 1 - {\hbar^2 \over m_B} \check{g}_{\bf p}  \check{R}_{\bf p} } \check{g}_{\bf p}   
\check{T}_{\bf p} \right] \vec{\Psi}_{\bf p}^{(F)}(0).
 \end{equation}
The boundary condition (\ref{b4}) together with (\ref{FEq2}) and (\ref{bc(0)}) form a self-consistent set for $z>0$. 

We assume at this point that the spin-orbit coupling energy scale is small compared to other relevant energies in the problem 
and keep the spin-orbit parameter $\alpha_R$ finite only where we otherwise would get a vanishing effect, i.e. in the 
spin-mixing terms. In all other quantities, we set $\alpha_R = 0$. This brings the reflection matrix to a simpler form 
proportional to the unit matrix,
\begin{equation}
\label{Rapr}
\check{R}_{\bf p} \approx \kappa \check{1} \equiv \kappa \hat{\tau}_0 \hat{\sigma}_0,
\end{equation}
where $\kappa$ is defined in (\ref{kappa}). In the $\alpha_R \to 0$ limit we have $\kappa = q/ \tanh{(q a)}$, 
where $q = \sqrt {{2m \over \hbar^2} (U_0 + {{\bf p}^2 \over 2m_B} - \varepsilon )}$. In the limit of a high barrier, 
$\kappa \approx q$.

We now consider the operator denominator in the boundary condition (\ref{bc(0)}). If the bulk superconductor is deep in 
the paired state, we may neglect the normal Green function and estimate the integrated Green function in (\ref{g}) as 
$\check{g} \approx f \hat{\tau}_x (i \hat{\sigma}_y )$, with 
\begin{equation}
\label{f}
f = {1 \over 2 \hbar v_S} {\Delta \over \sqrt{\Delta^2 + \xi_{\bf p}^2}}.
\end{equation}
For small $\xi_{\bf p} = p^2/(2m_S) - \mu_S$, it can be estimated as $f \approx 1/(2\hbar v_S)$, where $v_S$ is 
the Fermi velocity in the superconductor.  Therefore, the term in question from (\ref{bc(0)}) becomes
\begin{equation}
\label{renorm}
{\hbar^2 \over m_B} {1 \over 1 - {\hbar^2 \over m_B} \check{g}_{\bf p}  \check{R}_{\bf p}}  \check{g}_{\bf p} \approx 
{1 \over \kappa} {\alpha \over 1 + \alpha^2} \left[ \hat{\tau}_x (i \hat{\sigma}_y) - \alpha \check{1} \right],
\end{equation}
where $\alpha = \kappa f /m_B$.  This parameter can be estimated as $\alpha \sim \sqrt{m_S U_0/m_B E_F^{(S)}}$.
Note that the last term in (\ref{renorm}) is uninteresting because it does not include any off-diagonal contributions in 
Nambu space. The term slightly renormalizes the boundary conditions for the transverse wave function in the ferromagnet. 
It can therefore be safely dropped.
 
 Incorporating the right side of the boundary condition (\ref{bc(0)}) into the Schr\"{o}dinger equation for the
 ferromagnet (\ref{FEq2}) we obtain
 \begin{equation}
 \label{FP}
 \check{G}^{(F)}_{\bf p}\!\!{\phantom{|}}^{-1} \circ \vec{\Psi}_{\bf p}^{(F)}(z) = {-\hbar^2 \alpha  \delta(z) \over 2 f \kappa m_B (1 + \alpha^2)}
 \check{T}_{\bf p}  \check{g}_{\bf p}  \check{T}_{\bf p} \vec{\Psi}_{\bf p}^{(F)}(0).
 \end{equation}
 This is the Bogoliubov-de Gennes equation for the normal region with superconducting correlations introduced via the 
 boundary term that carries information about all Andreev processes. 
 
 \subsection{Proximity-induced superconducting gap}
 
 In order to calculate the superconducting gap induced in the ferromagnet, we assume that its magnetization is unaffected by the proximity
 to the superconductor and fix it instead of calculating it self-consistently. We also focus on a particular sub-band with a given spin polarization and  ignore intraband scattering (which is justified only in the clean case). Furthermore, we assume that the ferromagnetic layer or wire is very thin in the $z$-direction and  approximate  the full solution as
 \begin{equation}
 \label{product}
 \psi_{\alpha,{\bf p}}^{(F)}(z) = \chi_\alpha({\bf m}) \psi^{\rm (tr)}(z) \phi({\bf p}),
 \end{equation}
 where $\chi_\alpha({\bf m})$ is a spinor describing the magnetization direction that we wish to enforce and
 $ \psi^{\rm (tr)}(z)$ is the transverse envelope wave function that we approximate as a solution of the free one-dimensional Sch{\"o}dinger 
 equation, 
\begin{equation}
\label{tr}
\left( \partial^2_z  + {2 m_F \varepsilon_{\rm tr} \over \hbar^2} \right) \psi^{\rm (tr)}(z)=0,
\end{equation}
with the unusual boundary conditions,
\beq
{ m_B \over m_F} \partial_z \psi^{\rm (tr)}(0)  =  \kappa \psi^{\rm (tr)}(0),\qquad  \psi^{\rm (tr)}(d)  =0.
\eeq
We then introduce solution (\ref{product}) into (\ref{FP}), multiply it by $\psi^{\rm (tr)*}(z)$, average it over the transverse 
direction, and extract the relevant spin component 
(a word caution here is that the standard convention for the Gor'kov Green function taken proportional to 
$i\hat{\sigma}_y$ is basis-dependent and assumes spin quantization along the $z$-axis). 

For the ferromagnetic state polarized along $z$, we obtain the Bogoliubov-de Gennes equations in the familiar form,
\begin{equation}
\label{BdGp+ip}
\left[ {{\bf p}^2 \over 2m_F} - \left( \mu_F - \varepsilon_{\rm tr} \right) \right] \phi_{\bf p} 
+ E_g \left(p_y + i p_x \right) \phi_{-{\bf p}}^* = 0,
\end{equation}
where the proximity-induced gap or so-called minigap is
\begin{equation}
\label{Eg}
E_g = {f \hbar^4 \over p m_B^2} \kappa_t \delta\kappa_t \left|\psi^{\rm (tr)}(0)\right|^2.
\end{equation}
The boundary value of the transverse wave function can be obtained from (\ref{tr}) and reads
 \begin{equation}
 \label{tr(0)}
 \left|\psi^{\rm (tr)}(0)\right|^2 = {2 \over d} {1 \over 1 + \xi^2 + \xi/(k_{\rm tr} d)},
 \end{equation}
where $\xi = m_F q/(m_B k_{\rm tr})$ and $k_{\rm tr}$ is a solution to the eigenvalue problem, 
$\tan{(k_{\rm tr} d)}/(k_{\rm tr} d) = - m_B / (m_F qd)$, which determines the spectrum, 
$\varepsilon_{\rm tr} = \hbar^2 k_{\rm tr}^2/(2m_B)$.  
We see from Eq.~(\ref{Eg}) that the proximity-induced minigap is extremely sensitive to the actual width of the normal
region: the smaller the width, the more frequent Andreev scattering processes, and the larger the minigap.\cite{ProxEff2}

\section{resistance peak due to Superconducting fluctuations}
\label{sec:SCfluc}
We argued in Sec. \ref{meanfree} that the W electrodes have a strong impact on the transport properties
through the Co wire. We reiterate the experimental fact that when an extra W strip is deposited on top of the Co wire [see Fig.
\ref{fig:sys}(c)], the normal state resistance of the wire increases by nearly 50\%. This indicates that when a
W electrode is deposited onto a nanowire they modify the wire in its vicinity and provide a major source 
of resistance. A faithful model of the Co wire may then be such that the current $i_w$ goes through
the two W voltage electrodes. If this view is adopted this creates four W-Co interfaces, each one similar to 
the interface considered in Sec. \ref{andreev}. With this geometry electron transport through the 
wire is expected to be strongly modified by Andreev physics at the W-Co interfaces. 

We now focus on the resistance peak observed near the superconducting transition temperature of the W
electrodes. This peak is very reminiscent of similar anomalous peaks 
studied in the context of $c$-axis transport in cuprate superconductors\cite{SCfluc0,SCfluc1,SCfluc2}, magnetoresistance
in dirty films\cite{SCfluc3}, as well as magnetoresistance in granular electronic systems\cite{SCfluc4}. In all these cases,
the anomalous peak has been explained using the phenomenon of superconducting fluctuations.\cite{varlarrev,varlamovetal}
This phenomenon is concerned with fluctuating Cooper pairs that form while the system is in the normal state, either 
just above the transition temperature $T_c$ or the critical magnetic field $H_c$. For the $H=0$ case,
the appearance of Cooper pairs above $T_c$ opens up a new channel for charge transport. Indeed, these
fluctuating Cooper pairs can be treated as carriers of charge 2$e$ with a lifetime given by $\tau_{\rm GL}
\sim\hbar/k_B(T-T_c)$. This leads to a contribution to the conductivity known as the Aslamazov-Larkin (AL)
contribution or paraconductivity, and gives a positive correction to the Drude conductivity. 
There is also an indirect correction known as the density of states (DOS) contribution. 
One of the important consequences of these fluctuating 
Cooper pairs is the decrease in single-particle DOS near the Fermi level. The idea is that  if some electrons 
are involved in pairing they can not simultaneously participate in single-electron transport. The DOS contribution
therefore gives a negative correction to the Drude conductivity. 


As we approach $T_c$ of the W electrodes, Cooper pair fluctuations grow inside the W electrodes. 
By virtue of the Rashba spin-orbit coupling at the W-Co interfaces (due to breaking of mirror
symmetry in the longitudinal direction at the interface) singlet Cooper pairs inside W are converted and triplet 
fluctuations permeate into the Co wire. AL correction to the conductivity arises due to the transport of these fluctuating
Cooper pairs through the wire. However, Cooper pair tunneling across an interface is strongly suppressed because
it is a higher-order process in both tunneling and the singlet-triplet conversion rate, which 
requires a spin flip process enabled only by weak interfacial spin-orbit coupling. This can be seen in (\ref{T}) where the off-diagonal elements of the transmission matrix are suppressed by the factor $\de\ka_t$, which is proportional to $\al_R$. 
The single electron tunneling and the DOS correction, on the other hand, are lower order tunneling processes that require no 
spin flipping. We therefore qualitatively expect the DOS correction to be parametrically
larger than the AL contribution, thus giving a negative overall correction to the Drude conductivity. This clearly can
explain the anomalous upturn in the resistance as a function of temperature.

When the Co wire is replaced by a gold wire, the situation is modified. Since gold is not ferromagnetic, $s$-wave
pairing correlations are expected to survive over a much longer distance inside the wire.  In this case, Cooper pair tunneling through the
W-Au interface is much more transparent since it does not require spin flipping. Therefore, Cooper pairs
are more efficient at transporting charge in this case, and the AL contribution to the conductivity is expected to
play a much bigger role than with a ferromagnetic wire. Here, we purport that the absence of the resistance peak
in the Au wire can be explained if indeed the AL contribution to the conductivity is parametrically larger than the 
DOS contribution, thus giving an overall positive correction to the conductivity.

Near the superconducting transition temperature (and at $H=0$) and in dimensions at or below two, both AL
and DOS contributions show either algebraic or logarithmic divergent behavior as a function of the 
distance from the transition point. Denoting the divergent part of the AL corrections by function
$f_{\rm AL}(T-T_c)$ and the divergent part of the DOS correction by $f_{\rm DOS}(T-T_c)$ (and ignoring the Maki-Thompson
contribution),
the total superconducting fluctuation correction to the normal resistance $R_0$ can be schematically written as
\beq
\label{desig}
\frac{\de R(T)}{R_0}=A_{\rm DOS}f_{\rm DOS}(T-T_c) - A_{\rm AL}f_{\rm AL}(T-T_c) ,
\eeq
where, $A_{\rm AL}$ and $A_{\rm DOS}$ are positive pre-factors. As we will show in Sec. \ref{micro}, 
\beq
f_{\rm AL}\sim\frac{1}{(T/T_c-1)^4},
\eeq
while
\beq
f_{\rm DOS}\sim\frac{1}{\sqrt{T/T_c-1}},
\eeq
indicating that the AL contribution diverges much more strongly than the DOS term as $T\to T_c^+$.
Therefore, unless the pre-factor $A_{\rm AL}$ is 
exceedingly small in comparison to $A_{\rm DOS}$ the DOS contribution is always parametrically smaller than 
the AL contribution and the conductivity correction is positive. This leads to a monotonic decrease in 
the resistance as the system approaches the transition, and no peak will result. However, if we 
have a situation where the AL pre-factor is greatly suppressed, i.e. $A_{\rm AL}\ll A_{\rm DOS}$, the DOS contributions 
may become parametrically larger than the AL counterpart for $T$ sufficiently, but not too close to $T_c^+$. In this case, 
the resistivity can display an anomalous upturn before the AL contribution eventually takes over and the resistivity makes 
a precipitous fall as $T\to T_c^+$. 

\subsection{Microscopic theory}
\label{micro}
\subsubsection{Effective dimensionality of a tungsten electrode for fluctuation analysis}
The superconducting fluctuation physics depends critically on the dimensionality of the system, 
with the general trend being that the lower the dimensionality, the more pronounced and singular the fluctuation 
effects are. However, one should exercise care in determining the effective dimensionality of a system, as this notion 
depends on a particular effect that is being studied. For example, the system may be three-dimensional in terms of 
single-electron diffusion physics, but fall into the category of one-dimensional superconductors when it comes to the 
fluctuation analysis. We believe that such is the case with the W electrodes that host superconductivity 
and most of the fluctuation physics in the experiment under study involving  $L_{\perp}^{(1)}\sim$ 250nm 
wide and  $L_{\perp}^{(2)}\sim$100nm thick W strips. 

The finiteness of the transverse directions implies that whenever we have an integral over a three-dimensional 
momentum, its transverse part must be replaced with a sum over quantized modes,
$$
\int \frac{dq_\perp}{2 \pi \hbar} f(q_\perp) \longrightarrow {1 \over L_{\perp}} \sum\limits_{n_\perp} 
f \left( { 2 \pi \hbar n_\perp \over L_{\perp} } \right).
$$
For superconducting fluctuation analysis, the relevant function of interest  is the fluctuation propagator, which appears 
in the combination
\begin{equation}
\label{flsum}
{1 \over L_{\perp}} \sum\limits_{n_\perp} \left[ D \left(2 \pi \hbar n_\perp \over L_{\perp} \right)^2 + Dq_{||}^2 
+ \tau_{\rm GL}^{-1} \right]^{-1},
\end{equation}
where $D = v_{\rm F}^2 \tau/3$ is the diffusion coefficient and the Ginzburg-Landau relaxation time,
\begin{equation}
\label{tGL}
\tau_{\rm GL} = {\pi \over 8} {\hbar \over T - T_{\rm c}},
\end{equation}
tunes the proximity to the transition temperature $T_{\rm c}$. Hereafter, we will focus on a superconducting 
electrode and assume that we are dealing there with a disordered superconductor.

\begin{center}
\begin{figure}[t]
\includegraphics*[scale=0.6]{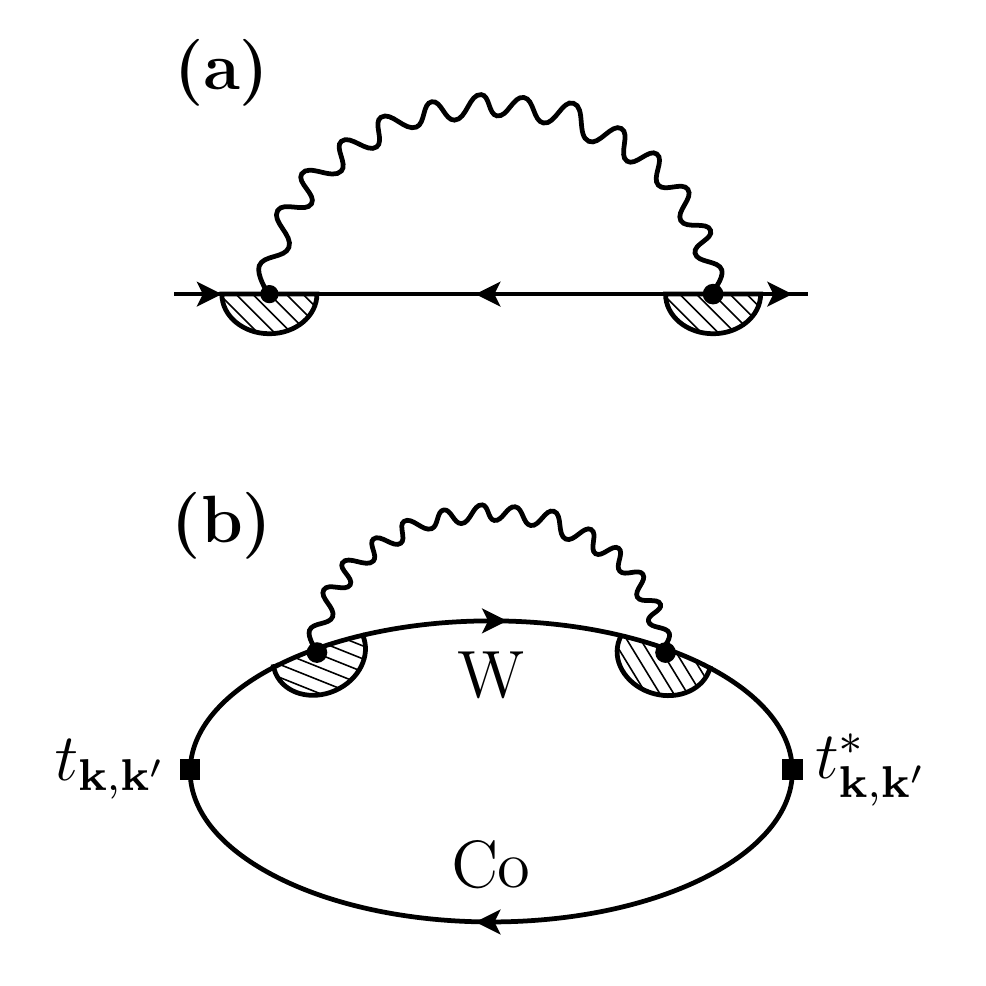}
\caption{\label{fig:DOS_Diag} (a) Self-energy diagram describing a correction to the single-electron Green
function. The wavy line corresponds to the superconducting fluctuation propagator, and the shaded vertices
represent Cooperon vertices. (b) A diagram corresponding to the lowest-order correction to the transport
kernel due to superconducting fluctuations in the tungsten electrode. As labeled in the figure, top (bottom) solidline represents the electron propagator for tungsten (cobalt).}
\end{figure}
\end{center}
The question of whether or not a particular dimension is important reduces to the comparison of the first and last 
term in the square brackets in (\ref{flsum}). If the former is much larger than the latter for any $n_\perp \ne 0$ 
the corresponding dimension is unimportant and an effective reduction of dimensionality occurs. 
One can define the following characteristic temperature scale (to be compared with $T-T_{\rm c}$) as follows
\begin{equation}
\label{Tperp}
T_{\perp}(L_\perp) = {1 \over k_{\rm B}} \left( {\pi^3 \over 6} \right) \left( {l \over L_\perp} \right) 
\left( { \hbar v_{\rm F} \over L_\perp} \right),
\end{equation}
where $v_{\rm F}$ is the Fermi velocity, $l = v_{\rm F} \tau$ is the electron mean-free path and we have restored 
the Boltzmann constant $k_{\rm B}$ and the Planck constant $\hbar$. 

For W, the Fermi velocity can be estimated as $v_{\rm F} \sim 0.5 \cdot 10^6$\ m/s and taking the largest 
of the two transverse dimensions $L_\perp \sim 250$\ nm, we find
$$
T_{\perp}^W(250\mbox{nm}) \sim  \left( {l \over L_\perp} \right) 80 \mbox{ K}.
$$
If we assume that the mean-free path in the amorphous W strips is of order a nanometer the corresponding 
temperatures scale becomes of order $T_{\perp}^W(250\mbox{nm}) \sim 0.3$-$1$K, which is very reasonable 
and implies that as soon as we approach the superconducting transition with $(T - T_{\rm c}) \ll T_{\rm c} \sim 4.4$-$5$K, 
we may view the electrode as a one-dimensional superconductor. 

\subsubsection{Fluctuation correction to the density of states}
\label{sec:dos}
We now analyze the DOS fluctuation physics on the basis of the standard diagrammatic perturbation theory. 
The Cooper-channel correction to the electronic DOS is given by
\begin{equation}
\label{dnu}
\delta \nu (\epsilon) = - {1 \over \pi} \int {d^d p \over (2 \pi)^d} {\rm Im}\, \left\{ {\cal G}^2(i\varepsilon_m, {\bf p} ) 
\Sigma (i\varepsilon_m, {\bf p} ) \right\}\phantom{|}_{i\varepsilon_m \to \epsilon},
\end{equation}
where $\Sigma (i\varepsilon_m, {\bf p} )$ is the self-energy  described by the diagram in Fig. \ref{fig:DOS_Diag}(a).
It reads
\begin{multline}
\Sigma (i\varepsilon_m, {\bf p} ) = - T \sum_{\Omega_n}
\int{d^d q \over (2 \pi)^d} {\cal G}(i \Omega_n - i\varepsilon_m, {\bf q} - {\bf p} )\\ 
\times {\cal C}^2 (\varepsilon_m, \Omega_n - \varepsilon_m;{\bf q}) {\cal L} (\Omega_n, {\bf q}),
\end{multline}
where
\begin{equation}
\label{G}
{\cal G} (i\varepsilon_m, {\bf p} ) = {1 \over i \tilde{\varepsilon}_m - \xi_{\bf p}}
\end{equation}
is the Matsubara Green function with $\xi_{\bf p} = v_{\rm F} (p - p_{\rm F})$ and 
$\tilde{\varepsilon}_m= \varepsilon_m + {\rm sgn}\,(\varepsilon_m)/(2\tau)$ and $\tau$ is the scattering time,
\begin{equation}
\label{C}
{\cal C} (\varepsilon_m, \Omega_n - \varepsilon_m;{\bf q})= {1\over \tau} 
{\theta\left[ \varepsilon_m (\varepsilon_m - \Omega_n) \right] \over Dq^2 + \gamma_{\rm s} + \left| 2\varepsilon_m 
- \Omega_n\right|}
\end{equation}
is the Cooperon vertex, where we included pair-breaking scattering rate $\gamma_{\rm s} = 2/\tau_{\rm s}$
and $\theta(\cdot)$ is the standard Heaviside step function. In the vicinity of the transition point 
the fluctuation propagator reads
\begin{equation}
\label{L}
{\cal L} (\Omega_n, {\bf q})= {8 T_{\rm  c} \over \pi \nu_0} 
\left[ Dq^2 + \tau_{\rm GL}^{-1} + |\Omega_n| \right]^{-1},
\end{equation}
where $\tau_{\rm GL}= {\pi \over 8} {\hbar \over T - T_{\rm c}}$ and $\nu_0=mp_F/2\pi^2$ is the bare DOS
for a single spin projection.
As per the usual convention, $\Omega_n = 2\pi n T$ denotes the bosonic Matsubara frequency and 
$\varepsilon_m = (2 m + 1) \pi T$ is the fermionic Matsubara frequency.  Finally, note that the physical quantity 
of interest, the DOS, must be analytically continued from the discrete set of Matsubara frequencies to 
the continuum of real energies, as labelled by the symbol ${i\varepsilon_m \to \epsilon}$ in (\ref{dnu}). 

Since the Tungsten electrodes are amorphous, we can safely assume that they are strongly disordered 
$s$-wave superconductors, where $T_{\rm c} \tau \ll 1$ (but of course we also assume that we are far from 
localization, that is $E_{\rm F} \tau \gg 1$). In this case, the three-Green-function block takes the especially simple form
\begin{multline}
\label{G3}
\nu_0 \int d\xi  {\cal G}(i \Omega_n - i\varepsilon_m, {\bf q} - {\bf p} ) {\cal G}^2(i\varepsilon_m, {\bf p} ) \\ = -
2 \pi i\nu_0 \tau^2 {\rm sgn}\, (\varepsilon_m),
\end{multline}
where we enforced the constraint $\varepsilon_m (\varepsilon_m - \Omega_n) > 0$ in (\ref{C}).
This leads to the following expression for the DOS
\begin{equation}
\label{dnu2}
\delta\nu (\epsilon) = {4 \over \pi^3} \int {d^d q \over (2 \pi)^d} \left( {\partial \over \partial \gamma_{\rm s}} \right)
{\rm Re}\, S^R({\bf q}, \epsilon),
\end{equation}
where $S^R({\bf q}, \epsilon) = S({\bf q}, i\varepsilon_m \to \varepsilon)$ represents the analytically-continued 
(retarded) Matsubara sum
\begin{align}
\label{S}
&S({\bf q}, \varepsilon_m>0) \\ &= \sum\limits_{n = -\infty}^{m} {1 \over 
\left( \frac{Dq^2+\tau^{-1}_{\rm GL}}{2\pi T_c} + |n| \right) 
\left( \frac{Dq^2 +\g_s}{2\pi T_c} + |2 m + 1 - n| \right)}.\nn
 \end{align}
This sum can be calculated exactly in terms of the digamma function, 
\beq
\psi(z) = -\gamma + \sum\limits_{n=0}^\infty \left[ \left(n+1\right)^{-1} -  \left(n+z\right)^{-1} \right],
\eeq
which is analytic everywhere except $z = 0, -1, -2, \ldots$. It is convenient to 
separate the sum (\ref{S}) into two pieces, $\sum_{n = -\infty}^{0} \cdots$ and $\sum_{n = 0}^{m} \cdots$.
For the first term analytic continuation reduces to replacing $\varepsilon_m \to -i \varepsilon$. For the
second term it becomes possible after noticing the ``reflection  property'', where $m-n$ can be replaced 
by $n - m$. This leaves the sum from $n=0$ to $n=m$ unchanged but the denominator ``positively-defined'' 
and ready for analytic continuation. 
Finally, the asymptotic form of DOS in the limit of $\left\{ \epsilon, \tau_{\rm GL}^{-1}, \gamma_{\rm s} \right\} \ll 
T \sim T_{\rm c}$ becomes
\begin{equation}
\label{Sres}
S^R({\bf q}, \epsilon)  = {(2 \pi T_{\rm c})^2 \over
\left(  Dq^2 + \tau_{\rm GL}^{-1} \right) \left(  2 Dq^2 + \gamma_s + \tau_{\rm GL}^{-1} - 2 i \epsilon \right)}.
\end{equation}

Let us now focus specifically on the correction to the DOS of electronic states with spin-up at the Fermi level, 
i.e. $\epsilon = 0$. The remaining elementary integrals in (\ref{dnu2}) can now be easily calculated and we find
\begin{equation}
\label{dnures}
\delta\nu_\uparrow (0,T) = -   {\nu_0 \over p_{\rm F}^2 A } \frac{\sqrt{6 \pi^7}}{128}
{T_{\rm c}^2 \tau^{-1/2} \over (T-T_{\rm c})^{5/2}} {1 + 2 \alpha(T) \over \alpha^3(T) \left[ 1 + \alpha(T) \right]^2},
\end{equation}
where
\begin{equation}
\label{alpha}
\alpha(T) = \sqrt{ {1 \over 2} \left[ 1 + {\pi \over 4} {1 \over \left( T-T_{\rm c} \right) \tau_{\rm s}} \right]}.
\end{equation}
Here, $A$ is the effective area of the fluctuating superconductor in the region where the fluctuations are studied
(we expect it to be related to the dimensions of the superconductor-nanowire contact and, consequently, expect 
$A$ to be smaller than the cross-sectional area of the wire). Notice that in the absence of pair breaking and/or
far from the transition, where $\left( T-T_{\rm c} \right) \tau_{\rm s} \gg 1$, (\ref{alpha}) reduces to a 
constant $\alpha = 1/ \sqrt{2}$ and the  DOS at the Fermi level acquires a very sharp temperature dependence
as follows:
\begin{equation}
\label{dnuas1}
\delta\nu_\uparrow (0,T) \propto - (T - T_{\rm c} )^{-5/2}, \mbox{ if } \left( T-T_{\rm c} \right) \tau_{\rm s} \gg 1.
\end{equation}
In the opposite regime of strong pair-breaking scattering or in the immediate vicinity of the transition, we find
\begin{equation}
\label{dnuas2}
\delta\nu_\uparrow (0,T) \propto - (T - T_{\rm c} )^{-1/2}, \mbox{ if } \left( T-T_{\rm c} \right) \tau_{\rm s} \ll 1.
\end{equation}

\subsubsection{Fluctuation correction to the contact resistance}

The result for the DOS at the Fermi level provides a useful insight into the physics near the transition 
and illustrates that, as a precursor to the global pairing transition, electrons are actively swept from the vicinity of 
the Fermi level due to the formation of the fluctuating Cooper pairs. However, the stand-alone quantity $\delta\nu(0,T)$ 
is not extremely useful for comparison with experiment, as it is actually not  directly measurable. What is measured in 
experiment is resistance, which is an integral quantity that includes excitations with different energies and that has 
contributions from both single-electron transport across the W-Co junction and pair transport. As found in Sec. \ref{andreev}, 
the latter is suppressed strongly far from the transition due to the very small pair tunneling probability, which requires 
spin-orbit-assisted spin flips. Single-electron transport on the other hand does not rely on any spin-orbit coupling, 
and spin-up electrons can tunnel freely from the electrodes into the ferromagnet. Hence, there is a regime close to 
the transition (but still far enough from the immediate vicinity, where the Cooper pairs eventually take over) where 
the single-electron tunneling dominates and is already strongly suppressed by fluctuations. This results in the upturn 
in the resistance as $T\to T_c^+$.

If we now focus entirely on single-electron transport and disregard the pair-breaking scattering, one would tend 
to conclude (incorrectly) that the tunneling resistance acquires a contribution proportional to that in (\ref{dnuas1}). 
This, however, is not so and the correction to the conductance is much weaker. 
This is because the tunneling electrons involve not only those precisely at the Fermi level, but 
all electron excitations within the shell of energies $E\sim (E_{\rm F} - T, E_{\rm F} +T)$. Hence, what matters is  
a redistribution of the DOS beyond the shell of energies that participate in transport (any redistribution 
within the shell is essentially not observable). 

All these phenomena are automatically accounted for within the standard diagrammatic theory of transport. Here, 
we proceed nearly in a one-to-one correspondence with Ref. \onlinecite{DiCastro}. The tunneling  current is 
derived from the transport kernel $Q(\omega_l)$ shown in Fig. \ref{fig:DOS_Diag}(b), 
\begin{equation}
\label{I(V)}
I(V) = - e {\rm Im}\, Q^R (\omega) \Bigl|_{\omega = eV},
\end{equation}
where $Q^R (\omega) $ is analytically-continued transport kernel and $V$ is the voltage across the tunneling contact. 
In the setup under consideration the Matsubara transport kernel reads explicitly 
\begin{eqnarray}
\label{Q}
Q (\omega_l)&=& -T^2\!\!\!\!\!\!\!\! \sum\limits_{\varepsilon_m; \Omega_n; {\bf k}, {\bf k}'} \left| t_{{\bf k},{\bf k}'} \right|^2 
{\cal G}_W^2 (\varepsilon_m, {\bf k} ) {\cal G}_{Co} (\varepsilon_m + \omega_l, {\bf k}' ) \nn\\
&\times& \int {dq \over 2\pi A} {\cal L} (\Omega_n, {\bf q}) {\cal C}^2 (\varepsilon_m, \Omega_n - \varepsilon_m;{\bf q}) \\
&\times&{\cal G}_W (\Omega_n - \varepsilon_m,{\bf q} - {\bf k} ),\nn
\end{eqnarray}
where $t_{{\bf k},{\bf k}'}$ is the tunneling amplitude between two momentum states, ${\cal G}_{W/Co} 
(\varepsilon_m, {\bf k})$ is the electron Green function for the W electrode/Co nanowire and all other quantities have 
been defined in Sec. \ref{sec:dos}. Following Ref. \onlinecite{DiCastro} we introduce a normal state resistance of 
the W-Co junction $R_{W/Co}^{(0)}$ and obtain
\begin{align}
\label{Q2}
Q (\omega_l)&= {\pi T^2 \over 2 e^2 R_{W/Co}^{(0)}}  \sum\limits_{\varepsilon_m,\Omega_n} {\rm sgn} (\varepsilon_m)\, 
{\rm sgn} \left( {\varepsilon_m}  + \omega_l \right)\\
&\times\theta\left[ \varepsilon_m (\varepsilon_m - \Omega_n) \right]   \int {dq \over 2\pi A} {{\cal L} (\Omega_n, {\bf q}) \over
 \left(Dq^2 + |2\varepsilon_m - \Omega_n| \right)^2}.\nonumber
\end{align}
Repeating the dimensionality-independent summations as in Ref. \onlinecite{DiCastro} (note that the clean case\cite{DiCastro} 
becomes technically equivalent to the dirty case if we notice that the frequency-dependence of the three-Green-function block 
in the reference is identical to that in the two Cooperons that appear in (\ref{Q2})) and evaluating the remaining integral over momentum 
$q$ we obtain the non-linear $I$-$V$ dependence
\begin{eqnarray}
\label{I(V)res}
\nonumber
I(V) = &-& {T \over 8 \pi^2 e } \sqrt{3 \pi \over 2} {1 \over R_{W/Co}^{(0)} \nu_0 v_{\rm F} A} 
{\rm Im}\, \psi'\left[ {1 \over 2} - {ieV \over 2\pi T} \right]\\
&\times& {1 \over \sqrt{(T-T_{\rm c}) \tau}},
\end{eqnarray}
where $\psi(\cdot)$ is the logarithmic derivative of the gamma-function. Now focusing on the linear response regime, 
${dI \over dV}|_{V=0}$, we find the leading fluctuation correction to the W-Co contact resistance
as follows,
\begin{equation}
\label{Rres}
{\delta R_{W/Co}^{(DOS)} \over R_{W/Co}^{(0)}} =  {7 \zeta(3) \over 4 \pi} \sqrt{3\pi \over 2} 
{1 \over p_{\rm F}^2 A} {1 \over \sqrt{(T - T_{\rm c})\tau}},
\end{equation}
where $\zeta(z)$ is the Riemann zeta-function and $\zeta(3) \approx 1.202$. The result in (\ref{Rres}) corresponds to a sharp 
upturn in the resistance upon approaching the superconducting transition of the electrode.
\begin{center}
\begin{figure}[t]
\includegraphics[bb=180 120 600 600,scale=0.33]{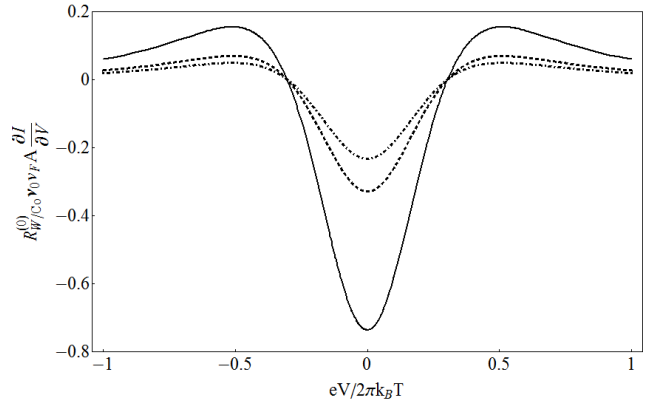}
\caption{\label{fig:DiffG} Correction to the differential conductance from superconducting fluctuations as a function of
dimensionless voltage $eV/2\pi T$ (see (\ref{I(V)res})). The three curves correspond to different temperatures:
$t:=(T-T_c)\tau=0.01$ for the solid line; $t=0.05$ for the dashed line; and $t=0.1$ for the dotted line.}
\end{figure}
\end{center}

\subsubsection{Estimate of tunneling resistance due to fluctuating pairs}

As emphasized throughout this section, the fluctuating Cooper pairs that appear in the electrodes do not aid with transport 
through the ferromagnetic nanowire initially because of their poor ability to tunnel into it. However, the experimental fact that 
the wire as long as a micrometer does become superconducting ensures that the pairs eventually are able to tunnel. Sec.~\ref{andreev} 
provides a microscopic picture of the Andreev reflection/boundary physics that presumably makes this possible. Therefore, we 
expect that as the temperature is tuned down to the closest vicinity of the transition, the Cooper pair tunneling takes over the 
effect of the suppression of the DOS and the upturn in the resistance crosses over to the downturn going through 
a peak, as observed in experiment. 

We notice here in passing that the height of the peak may provide a valuable insight into the competition of the two phenomena 
and may potentially become a means to measure the boundary spin-orbit coupling that is crucial for the proximity-induced 
$p$-wave superconductivity in the wire, the phenomenon of major interest. However, we leave the complicated microscopic 
theory of spin-orbit-assisted fluctuating pair tunneling for future studies and in this section only extract the leading temperature 
dependence of this AL type correction. It can be done on the basis of the Ambegaokar-Baratoff formula that 
yields\cite{varlarrev}
\begin{equation}
\label{ABAL}
\delta R^{(AL)}_{W/Co} \propto -\delta(\alpha_R) \int\limits_{-\infty}^\infty d\varepsilon {\left[\delta \nu (\varepsilon)\right]^2 
\over \cosh^2{\left(\varepsilon \over 2 T \right)}},
\end{equation}
where $\delta \nu (\varepsilon)$ is the suppression of the DOS obtained in Sec. \ref{sec:dos} and
we kept a small coefficient $\delta(\alpha_R)$ that includes spin-orbit suppression of the pair-tunneling as obtained in 
Sec.~\ref{andreev}.

Note that the integral of the DOS over all energies vanishes identically $\int_{-\infty}^\infty d \varepsilon 
\delta \nu (\varepsilon) =0$, (the physical interpretation being that the total electron density is conserved and the electrons 
can only be redistributed across different energies) and a non-zero correction to the resistance in the first order appears 
only due to the modulating function, $\cosh^{-2}{\left({\varepsilon \over 2T} \right)}$, and consequently is much weaker 
(see (\ref{Rres})) than that in the DOS (see (\ref{dnuas1})). For the pair tunneling, the situation is different as 
$\int_{-\infty}^\infty d \varepsilon \delta \nu^2 (\varepsilon)$ is not only non-zero, but is very singular function near the
transition. This singularity determines the scaling of the AL correction (\ref{ABAL}) as
\begin{equation}
\label{ABALres}
\delta R^{(AL)}_{W/Co}  \propto -{\delta(\alpha_R) \over \left(T - T_{\rm c}\right)^4}.
\end{equation}


\section{Topological superconductivity and Majorana fermions}
\label{TopSC}
\subsection{Relation to the experiment in Ref. \onlinecite{wangetal}}

The canonical Kitaev model\cite{kitaev}, which is the simplest prototype of a one-dimensional topological superconductor, 
involves one fermion species hopping in a one-dimensional chain and subject to a prescribed $p$-wave pairing field on 
the nearest-neighbor bonds. The end points of the chain host single Majorana excitations that are of key interest. 
One can consider ${\cal N}$ replicas of the Kitaev-Majorana model with different $p$-wave pairing fields for each species 
but with no mixing between them, i.e. no interband pairing. If ${\cal N}$ is even, the end Majoranas are unstable against 
various perturbations and generally hybridize into relatively uninteresting finite-energy boundary states. In contrast, if ${\cal N}$ 
is odd, the system can then host one Majorana zero-energy state at each end. Recently, theoretical works have addressed 
a multi-channel generalization of Majorana end states in quasi-one-dimensional structures with Rashba spin-orbit 
coupling.\cite{Q1DMF1,Q1DMF2,Q1DMF3,Q1DMF4}
The works show that the Majorana end states are realized in some parameter regime as long as an odd number of transverse 
subbands are occupied.

The well-known difficulty in realizing the regime where (\# of end Majoranas) mod 2 = 1 is of course related to the fact that the electrons 
have spin and consequently come in pairs. Therefore, according to Kitaev\cite{kitaev}, one necessarily, but not sufficiently, needs 
to break time-reversal symmetry in the one-dimensional superconductor to render it topological. The notable and fascinating proposals, 
where this regime was theoretically shown to be possible, include a heterostructure involving $s$-wave 
superconductor/spin-orbit-coupled wire/ferromagnet\cite{MF3,MF12}, a similar heterostructure without a ferromagnet but in a magnetic 
field\cite{MF4,MF10}, topological insulator/superconductor in a field\cite{MF1,MF13}, and half-metal/superconductor with spin-orbit coupling in the
latter\cite{MF11}.

If our interpretation of the experimental data in Ref. \onlinecite{wangetal} is correct, we see that this work potentially possesses all 
necessary ingredients for the realization of the Majorana end states; it has a ferromagnetic crystalline wire in which $p$-wave 
superconductivity has been induced. The ${\cal N}$-replica Kitaev-Majorana model discussed 
here may apply to a thin narrow Co film deposited on top of a three-dimensional W superconductor. The Co film should be
highly confined in the direction normal to the Co-W interface so that only one channel is occupied in that direction, and the width
of the strip should be smaller than the $p$-wave coherence length. Since Co is not a half-metal the question is whether 
a ferromagnetic wire, with both majority and minority carriers, can still host an odd number of end Majorana fermions. 
This, in principle, may be possible, as what we need is not necessarily to eliminate completely a spin component, but merely to 
make the two components different from each other such that the total number of occupied subbands is odd, i.e.
$\left({\cal N}_\uparrow + {\cal N}_\downarrow \right) {\rm mod}\, 2 = 1$. In the simplest model with no inter-subband mixing 
(considered here), this 
would imply topological superconductivity. However, we mention that a single-species $p$-wave superconductor, more akin to the
original proposal by Kitaev\cite{kitaev}, may be possible if the thin film Co wire is replaced by a similar film made of a 
half-metal such as CrO$_2$. Notably, experiments have recently shown that CrO$_2$ can accommodate long-ranged  
$p$-wave superconductivity when it is proximity-coupled to a conventional superconductor.\cite{SFJ4} 
Majorana bound states should then be realized at the ends of such a half-metal wire 
when odd number of transverse subbands are occupied.\cite{Q1DMF1}


\subsection{Realizing Majorana end states using a ferromagnetic semiconductor/ferromagnetic superconductor heterostructures}

Ref.~\onlinecite{wangetal} and the discussion above suggest an even simpler experimental setup for topological superconductivity.
Indeed, realizing a one-dimensional topological superconductor using spinful fermions requires lifting the double degeneracy imposed
by time-reversal symmetry. In principle, this can be achieved by proximity-inducing a Zeeman gap\cite{MF3} or by applying
an external magnetic field\cite{MF4,MF9}. An alternative approach to realizing a one-dimensional topological superconductor
is to deposit a ferromagnetic semiconductor wire on top of a ferromagnetic superconductor\cite{U1,U2,U3,U4}. A particularly
attractive candidate ferromagnetic semiconductor is europium oxide (EuO), which is known to possess nearly spin-polarized 
bands.\cite{euo1,euo2}  EuO becomes ferromagnetic below 70K under ambient pressure and the Curie temperature
is known to increase with pressure reaching 200K under $1.5\times 10^5$ atmospheres.\cite{euo3} Since its integration with
with Si and GaN\cite{euo2}, EuO has garnered much attention for its potential use in spintronic applications.
Ferromagnetic superconductors are materials that exhibit intrinsic coexistence of superconductivity and ferromagnetism where 
the same electrons are believed to be superconducting and ferromagnetic simultaneously. 
Perhaps the most well-known experimental realizations of ferromagnetic
superconductors are the Uranium-based compounds. Following the first experimental realization a decade ago there are now four 
such compounds\cite{U1,U2,U3,U4}, two of which exhibit the phenomena at ambient pressures\cite{U2,U3}. 
We propose that depositing URhGe electrodes on the EuO wire, in the arrangement shown in Fig. \ref{fig:sys}(a),
would be conceptually the simplest structure to realize 
Majorana fermions that would require neither topological insulators nor control over spin-orbit couplings. Application
of a magnetic field can further stabilize both the ferromagnetism and superconducting phase.

\section{Summary}
In this work, we study temperature-dependent transport properties of a ferromagnetic cobalt nanowire proximity-coupled
to superconducting electrodes. The work is largely motivated by a recent experiment,\cite{wangetal} in which 
long-ranged superconducting proximity effect was observed in single-crystal Co nanowires coupled to conventional superconductors. 
We focus particular attention to wires where the transition to superconductivity is preempted by a large and sharp 
resistance peak near the transition temperature of the electrodes. 
We explore the possibility of inducing spatially-odd $p$-wave correlations in the wire via Rashba spin-orbit 
coupling in the wire-superconductor interface. We then theoretically study the physics of an $s$-wave-superconductor-ferromagnet tunnel junction in the presence of an interfacial Rashba spin-orbit coupling, and derive an expression for the induced $p$-wave minigap in terms of the microscopic parameters of the contact.  

In the second half of the work, we show how the anomalous resistance peak can be explained within the
theory of superconducting fluctuation corrections to conductivity. In particular, we develop a microscopic theory for the
superconducting fluctuation corrections in the tungsten electrodes and discuss how these corrections may lead to corrections in the 
measured resistance. In the last part of the work, we discuss in detail the possible relevance of the experiment to topological superconductivity and propose a related hybrid system, which provides the simplest physical realization of topological superconductivity and localized Majorana modes.

We note in conclusion that the spin-orbit-assisted proximity-effect scenario considered here is different from the diffusive ferromagnet case, where the likely mechanism for proximity effect is believed to be the odd-frequency pairing.\cite{efetovprl,SFJrev} The main argument for the odd-frequency superconductivity is that any anisotropic pairing would decay fast into a disordered metal on a length-scale of order a mean-free-path, while the odd-frequency isotropic pairing is immune from non-magnetic static disorder. This result follows from the Usadel equation if we assume that the non-$s$-wave part of the disorder-averaged  condensate wave-function is small. Then upon linearization of the Usadel equation, further exponential decay of the disorder-averaged $p$-wave condensate wave-function inevitably follows. We note however that in certain geometries (e.g., if the mean-free path is much larger than the cross-section of the wire, but much smaller than its length), there is no reason to assume that the anisotropic component is small locally near the contacts and consequently the conventional derivation of the Usadel equations from the Eilenberger equations breaks down in these regions. With such assumption about the boundary conditions, the correlator of  $p$-wave condensate wave-functions is not expected to decay exponentially at $T=0$, in sharp contrast to the exponential decay of the disorder-averaged $p$-wave condensate wave-function.\cite{pwpe,msfl1,msfl2} These mesoscopic fluctuation effects will be considered in detail elsewhere.\cite{gt}

\section*{Acknowledgments}
V. G. would like to thank K. Efetov, S. Das Sarma, Y. Oreg, J. Paglione, G. Tkachov and A. Varlamov for discussions. S. T.
would like to thank J. Wang for sharing their experimental data, and J. R. Anderson for informative 
discussions on the cobalt band structure. S. T. would also like thank J. D. Sau for stimulating discussions at an early stage of the work.
S. T. was supported by DOE and V. G. by DOE-BES (DESC0001911).

\bibliographystyle{plain}
\bibliography{SFjunction}
\end{document}